\documentclass[aps,pra,amsmath,amsfonts,amssymb,twocolumn]{revtex4}
\usepackage[latin1]{inputenc}
\usepackage{graphicx}
\usepackage[english]{babel}
\usepackage[usenames,dvipsnames]{color}

\usepackage{amsfonts}
\usepackage{amsmath}
\usepackage{amssymb}

\batchmode

\usepackage{color}

\usepackage{soul}

\newcommand{\beq}{\begin{equation}}
\newcommand{\eeq}{\end{equation}}

\newcommand{\ket}[1]{|#1\rangle}
\newcommand{\bra}[1]{\langle #1|}

\makeatletter
\@ifundefined{textcolor}{}
{%
 \definecolor{BLACK}{gray}{0}
 \definecolor{WHITE}{gray}{1}
 \definecolor{RED}{rgb}{1,0,0}
 \definecolor{GREEN}{rgb}{0,1,0}
 \definecolor{BLUE}{rgb}{0,0,1}
 \definecolor{CYAN}{cmyk}{1,0,0,0}
 \definecolor{MAGENTA}{cmyk}{0,1,0,0}
 \definecolor{YELLOW}{cmyk}{0,0,1,0}
 }

\begin{document}

\title{Mach-Zehnder interferometer with quantum beamsplitters}

\author{N. Almeida
$^{1}$
}

\author{T. Werlang
$^{1}$
}

\author{D. Valente
$^{1}$
}
\email{valente.daniel@gmail.com}

\affiliation{
$^{1}$ 
Instituto de F\'isica, Universidade Federal de Mato Grosso, CEP 78060-900, Cuiab\'a, MT, Brazil
}

\begin{abstract}

A one-dimensional waveguide enables a single two-level emitter to route the propagation of a single photon, as to provide a quantum mirror or a quantum beamsplitter.
Here we present a fully-quantum Mach-Zehnder interferometer (QMZ) for single-photon pulses comprised of two quantum beamsplitters.
We theoretically show how nonlinearities of the QMZ due to photon-emitter detunings and to the spectral linewidth of the pulse contribute to the versatility of the device with respect to the classical-beamsplitters scenario.
We employ a quantum dynamics framework to obtain analytical expressions for the photodetection probabilities and prove, in the monochromatic regime, the equivalence with a transfer-matrix approach.

\end{abstract}


\maketitle
\section{Introduction}
Coherent control of single-photon emission, absorption and transport opens promising perspectives for quantum communication and information processing, since photons can act as flying qubits between distant atomic nodes \cite{cirac97,rempe,bus,qnet,baranger}.
Both real \cite{rempe,kimble,haroche} and artificial atoms, either semiconducting \cite{claudon,lodahl,senellart} or superconducting \cite{tsai,devoret,walraff.filipp}, have been experimentally investigated as single-photon emitters.
The engineered electromagnetic environments required for controlling matter-field couplings at the single-photon level are usually implemented with optical cavities.
Because cavities trap and spectrally modify emitted photons \cite{imamoglu, dv1}, alternatives that employ one-dimensional (1D) waveguides have also been investigated \cite{domokos,fan,kojima,chang.lukin,dvaa,dvaa2,lodahl,OL}.
This research line, often called waveguide quantum electrodynamics (waveguide QED) \cite{ballestero14,baranger,rabl,cirac,lodahl17}, is designed to offer single-photon control for propagating light and can be implemented in diverse experimental platforms, ranging from nanophotonics \cite{claudon, lodahl, kimble.chang, rauschenbeutel, lodahl15, lodahl17,skolnick} to circuit quantum electrodynamics \cite{astafiev,wallraff,painter}.

Waveguide QED scenarios are particularly suitable for exploring interference effects on single-photon transport.
A key example is the full reflection of a single photon propagating in a 1D waveguide coupled to a single two-level system (TLS), due to a destructive interference at resonance \cite{domokos,fan,astafiev,lodahl2}.
The waveguide-TLS system plays the role of a quantum mirror, in this case.
A Fabry-Perot cavity made of two quantum mirrors can provide nonlinear and nonreciprocal photonic transport, as recently shown theoretically \cite{QFP,ballestero16,dudu,ff} and experimentally \cite{prldiode}.
The waveguide-TLS system can also act as a quantum beamsplitter, creating a superposition state of partially reflected and transmitted photon \cite{domokos,fan}.
Nonlinearity of a quantum beamsplitter has been evidenced for the scattering of two photons in a Hong-Ou-Mandel setup, where photon-photon correlations are mediated by the two-level emitter \cite{scarani}.
Interestingly, even the scattering of a single-photon pulse on a quantum beamsplitter shows a  clear nonlinear behavior as a function of the pulse linewidth, which is quantitatively different from the scattering of a low-intensity coherent (semiclassical) pulse \cite{domokos}.

Here, we theoretically investigate the use of two concatenated quantum beamsplitters to form a quantum Mach-Zehnder interferometer (QMZ) for a single-photon pulse.
We employ a dynamical approach for describing the pulse scattering.
This allows us to find analytical expressions for the scattered quantum state in real-space representation, as well as for the photodetection probabilities, as functions of the detunings and the pulse linewidth.
We look for the set of parameters for the QMZ to match its version with classical beamsplitters, as well as for the set where nonclassical signatures take place.
Both linear and nonlinear regimes of our QMZ are addressed.
We develop a transfer-matrix approach to be compared with the dynamical approach, looking for an equivalence in the monochromatic (linear) regime.
Our results reveal useful resources for adjustable elementary interferometers, made of single two-level emitters in 1D waveguides, realizable in state-of-the-art nanophotonic and superconducting circuit platforms.

\section{Model}
%
%
\begin{figure}[!htb]
\centering
\includegraphics[width=1.0\linewidth]{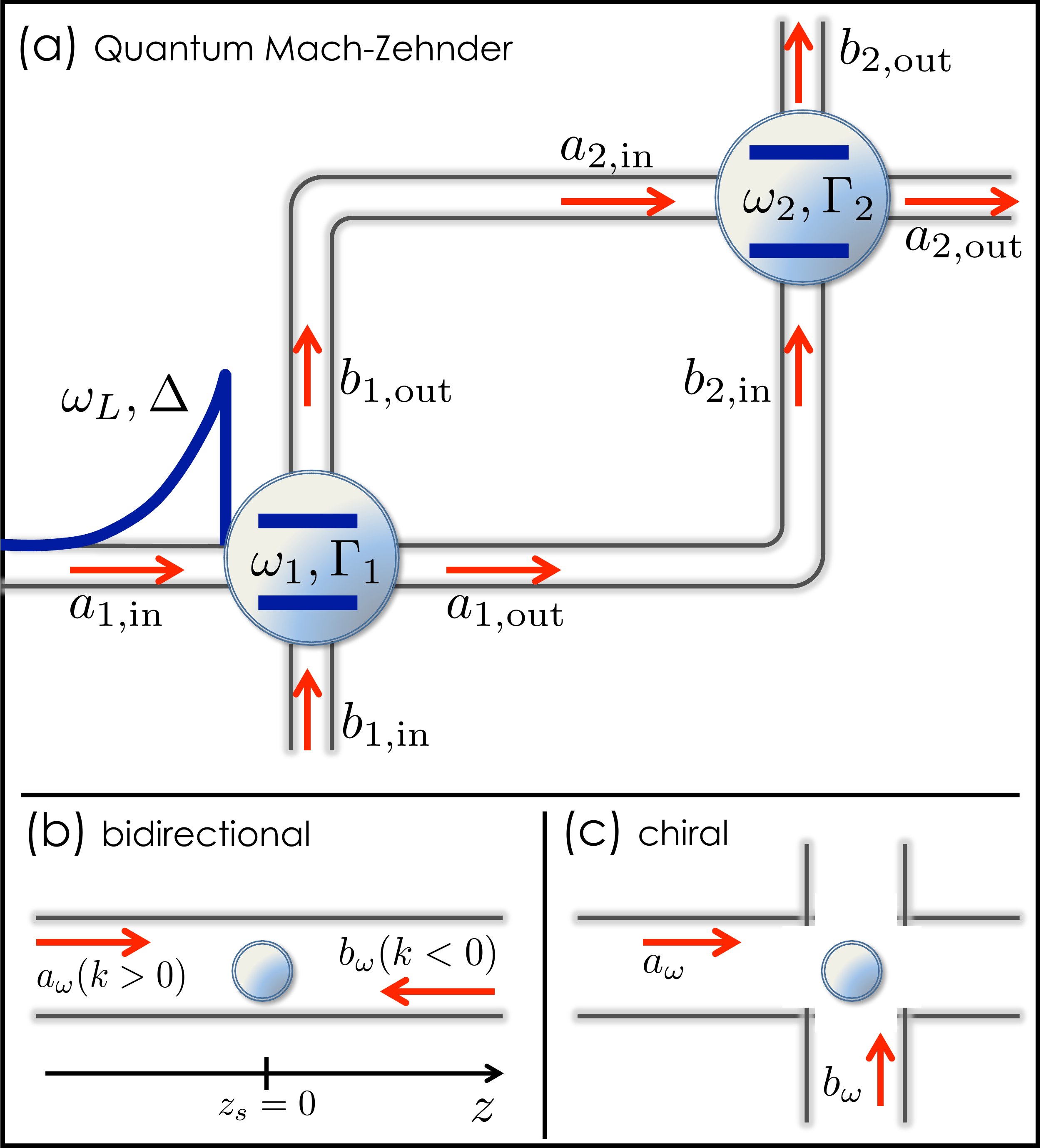}
\caption{(Color online)
{\bf Mach-Zehnder interferometer with quantum beamsplitters (QMZ).}
(a) A right-propagating single-photon pulse of central frequency $\omega_L$ and spectral linewidth $\Delta$ (exponential-shaped blue line) is sent through channel $a_{1,\mathrm{in}}$ of a waveguide.
The photon interacts with TLS $1$, of transition frequency $\omega_1$ and decay rate $\Gamma_1$.
The scattered (or absorbed and reemitted) photon goes through channels $a_{1,\mathrm{out}}$ and $b_{1,\mathrm{out}}$ that are connected, via waveguides, to channels $a_{2,\mathrm{in}}$ and $b_{2,\mathrm{in}}$.
After interaction with TLS $2$, of frequency $\omega_2$ and decay rate $\Gamma_2$, the photon comes out of the QMZ from channels $a_{2,\mathrm{out}}$ and $b_{2,\mathrm{out}}$.
Decay rates $\Gamma_{1,2}$ both describe spontaneous emission exclusively inside the waveguides.
(b) single bidirectional 1D waveguide. (c) pair of chiral 1D waveguides. 
Both (b) and (c) are potentially employable in the realization of a QMZ.
We fix notation for case (b).
}
\label{fig1}
\end{figure}
We model the scenario sketched in Fig.\ref{fig1}(a), where two TLSs shall act as quantum beamsplitters to form a QMZ for single-photon pulses.
We derive the dynamics of our QMZ by decomposing it in two successive and concatenated scattering events.
Each event is modeled as an interaction of a single-photon pulse with a single TLS coupled to a 1D waveguide.
In the case of a bidirectional waveguide, each propagation direction is associated with a different mode, labeled $a_\omega$ for the forwards and $b_{\omega}$ for the backwards scattering, as illustrated in Fig.\ref{fig1}(b).
The bidirectional 1D waveguide can alternatively be replaced by a pair of chiral waveguides \cite{rauschenbeutel, skolnick, lodahl15,ballestero16, lodahl17,lodahlchiralnature17}.
The chiral waveguide of propagating modes $a_\omega$ and the one of modes $b_\omega$ can be both coupled to the TLS and set in an orthogonal geometry, for instance, as illustrated in Fig.\ref{fig1}(c).
Below, we choose the bidirectional-waveguide perspective, Fig.\ref{fig1}(b), to fix notation in our modeling.

The dynamics of the composite waveguide-TLS system is unitary, governed by the total Hamiltonian 
$H = H_{\mathrm{TLS}}+H_{\mathrm{int}}+H_{\mathrm{field}}$, where
$H_{\mathrm{TLS}} = \hbar \omega_{0} \sigma_{+}\sigma_{-}$
is the Hamiltonian of the TLS, 
$\sigma_{-} = \sigma_{+}^\dagger = \ket{g}\bra{e}$, 
and the TLS ground (resp. excited) state is denoted by $\ket{g}$ (resp. $\ket{e}$).
The Hamiltonian of the 1D free field modes (see Fig.\ref{fig1}(b)) that propagate forwards $a_\omega$ and backwards $b_\omega$, with frequency $\omega$, inside the waveguide reads
$H_{\mathrm{field}} = \sum_{\omega} \hbar \omega [a^\dagger_\omega a_\omega + b^\dagger_\omega b_\omega] $.
The TLS-field interaction Hamiltonian, in the dipole and rotating-wave approximations, is given by 
\cite{domokos}
\beq
H_{\mathrm{int}} = 
\sum_\omega -i \hbar g 
[\sigma_{+} (a_\omega e^{+i k_\omega z_s}+ b_\omega e^{-i k_\omega z_s}) - \mbox{H.c.}],
\eeq
in which $z_s$ is the position of the TLS (set to $z_s = 0$), 
$k_\omega = \omega/c$ is the wavevector modulus, $c$ is the group velocity inside the waveguide, $g$ is the coupling rate and H.c. denotes the Hermitian conjugate.

We are interested in the single-excitation subspace, as described by the normalized pure state of the global TLS-plus-field system, 
\beq
\ket{\xi(t)} = \psi(t) \ket{e,0} +
\sum_\omega [\phi^{(a)}_\omega(t) a^\dagger_\omega + \phi^{(b)}_\omega(t) b^\dagger_\omega ] \ket{g,0},
\eeq
where $\ket{0}$ is the vacuum state of the field.
The excited-state population of the TLS is $|\psi(t)|^2$.
The real-space representation for the field reads \cite{kojima,dvaa, dvaa2,OL} 
\beq
\phi^{(a)}(z,t) = \sum_{\omega} \phi^{(a)}_{\omega}(t) e^{ik_{\omega}z} 
\eeq
for the $a_\omega$ modes.
A continuum of frequencies is assumed in the 1D environment, 
$\sum_\omega \rightarrow \int d\omega \rho_{\mathrm{1D}}$, 
so the flat spectral density of guided modes is named $\rho_{\mathrm{1D}}$.
$\phi^{(a)}(z,t)$ gives the probability amplitude that the photon is found at position $z$, at time $t$, propagating forwards.
For the  $b_\omega$ modes, one substitutes $k_\omega$ for $-k_\omega$ as the photon propagates backwards.
The photodetection probabilities read
\beq
p^{(a),(b)}(t)=\frac{1}{2\pi \rho_{\mathrm{1D}} c}\int_{-\infty}^{\infty} |\phi^{(a),(b)}(z,t)|^2 dz,
\label{pab}
\eeq
showing that we need to solve the dynamics for the field amplitudes.

\section{Results}
\subsection{Dynamics of single-photon pulse scattering on a single TLS in a bidirectional 1D waveguide}
We analytically solve the Schr\"odinger equation $i\hbar \partial_t \ket{\xi(t)} = H\ket{\xi(t)}$ to find the composite field-TLS dynamics.
The 1D continuum of frequency modes imposes a decay rate to the TLS, 
$\Gamma_{\mathrm{1D}} = 4\pi g^2 \rho_{\mathrm{1D}}$, 
that arises from the equations after a Wigner-Weisskopf approximation \cite{domokos,kojima,OL}.
Here, the decay of the TLS causes spontaneous emission only in the guided modes.
For a general initial state $\ket{\xi(0)}$, the excited-state amplitude dynamics is 
$
\psi(t) = \psi_{\mathrm{em}}(t) + \psi^{(a)}_{\mathrm{exc}}(t)+\psi^{(b)}_{\mathrm{exc}}(t)
$,
where the emission term reads
$
\psi_{\mathrm{em}}(t) 
= 
\psi(0) \exp\left[  -\left( \Gamma_{\mathrm{1D}}/2 + i\omega_0 \right) t \right]
$.
The excitation terms read
\beq
\psi^{(a)}_{\mathrm{exc}}(t) = 
-g \int_0^t \phi^{(a)}(0,t')  
 \ e^{-\left(\frac{\Gamma_{\mathrm{1D}}}{2} + i\omega_0 \right)(t-t')} dt'
\label{psi}
\eeq
and analogously for $\psi^{(b)}_{\mathrm{exc}}(t)$.
The initial packet condition, namely $\phi^{(a),(b)}(z,0)$, is applied by using that $\phi^{(a),(b)}(0,t')=\phi^{(a),(b)}(\mp ct',0)$,
which means that the TLS is driven at time $t'$ and position $z_s=0$ by the value of the initial photon packet at a distant position, $\mp ct'$.
Amplitudes $\phi^{(a),(b)}(z,0)$ set the states of the input modes $a_{\mathrm{in}}$ and $b_{\mathrm{in}}$ (see Fig.\ref{fig1}).

The general solution for the field amplitudes in real-space representation, $\phi^{(a)}(z,t)$ and $\phi^{(b)}(z,t)$, read
\begin{eqnarray}
\phi^{(a),(b)}(z,t) &=& \phi^{(a),(b)}(z\mp ct,0) \nonumber\\
&+& \beta \ \Theta(\pm z)\Theta(t\mp z/c) \ \psi(t\mp z/c),
\label{phia}
\end{eqnarray}
where $\beta = \sqrt{\Gamma_{\mathrm{1D}} \pi \rho_{\mathrm{1D}}}$ and $\Theta(z)$ is the Heaviside step function.
$\phi^{(a),(b)}(z,t)$ provide the state of the output modes $a_{\mathrm{out}}$ and $b_{\mathrm{out}}$ (see Fig.\ref{fig1}).
Eqs.(\ref{phia}) clearly reveal the interference between two amplitudes, one for the free propagation of the input photon and the other for the photon emitted by the TLS.
This interference is the central concept explored in the following sections.

\subsection{Single quantum beamsplitter}
\label{BS1}
The quantum beamsplitter functionality is detailed here.
Our purpose is not only to recover the results from Refs.\cite{domokos,fan}, but also to emphasize the main operation regimes that will be relevant to the following sections.
We choose an initial ground state, $\psi_1(0) = 0$, for the now labeled TLS $1$, in Eqs.(\ref{psi}) and (\ref{phia}).
Mode $b_{1,\mathrm{in}}$ starts in the vaccum state, $\phi^{(b)}_1(z,0) = 0$.
The input photon is prepared in channel $a_{1,\mathrm{in}}$ with a spontaneous emission profile,
\beq
\phi^{(a)}_1(z,0) = N \Theta(-z)e^{\left( \frac{\Delta}{2}+i\omega_L \right)  \frac{z}{c}}.
\label{initialpulse}
\eeq
$N=\sqrt{2\pi \rho_{\mathrm{1D}} \Delta}$ is a normalization factor.
The pulse is characterized by the spectral linewidth $\Delta$ and the central frequency $\omega_L$.
We set $\Gamma_{\mathrm{1D}} = \Gamma_1$ and $\omega_0 = \omega_1$ as the parameters of TLS $1$. 
We compute the photodetection probabilities, Eqs.(\ref{pab}), in the long time limit,  $t_\infty \gg \Gamma_1^{-1} + \Delta^{-1}$, in which TLS $1$ returns to its ground state.
We find analytical expressions for $p^{(a)}_1$ and $p^{(b)}_1$ as functions of the linewidth $\Delta$ and the detuning $\delta_1 = \omega_L - \omega_1$ (see Appendix). 

In Fig.(\ref{fig2}), we plot $p^{(a)}_1$ (dashed) and $p^{(b)}_1$ (full lines).
Fig.\ref{fig2}(a) shows the monochromatic regime, $\Delta \ll \Gamma_1$, where $p^{(b)}_1$ is a Lorentzian function of $\delta_1$ and $p^{(a)}_1  = 1 - p^{(b)}_1$.
At resonance, $\delta_1 = 0$, the TLS completely reflects the incoming photon, hence acting as a quantum mirror, as mentioned at the introduction.
Fig.\ref{fig2}(b) shows the nonlinear variation of the probabilities with respect to $\Delta$, at resonance (black) and off-resonance (red), as also shown in \cite{domokos}.
The balanced quantum beamsplitter condition, defined as $p^{(a)}_1 = p^{(b)}_1 = 1/2$, can be obtained at
\newline
\newline
{\bf(i)} $\delta_1 = \pm \Gamma_1/2$, for $\Delta \ll \Gamma_1$ (off-resonance monochromatic regime) and
\newline
\newline
{\bf (ii)} $\Delta = \Gamma_1$, for $\delta_1 = 0$ (resonant finite-linewidth regime).
\newline
\newline
Although configurations (i) and (ii) are not unique for the balanced condition, they bring more clarity to the analysis of the QMZ, performed in the following section.

\begin{figure}[!htb]
\centering
\includegraphics[width=1.0\linewidth]{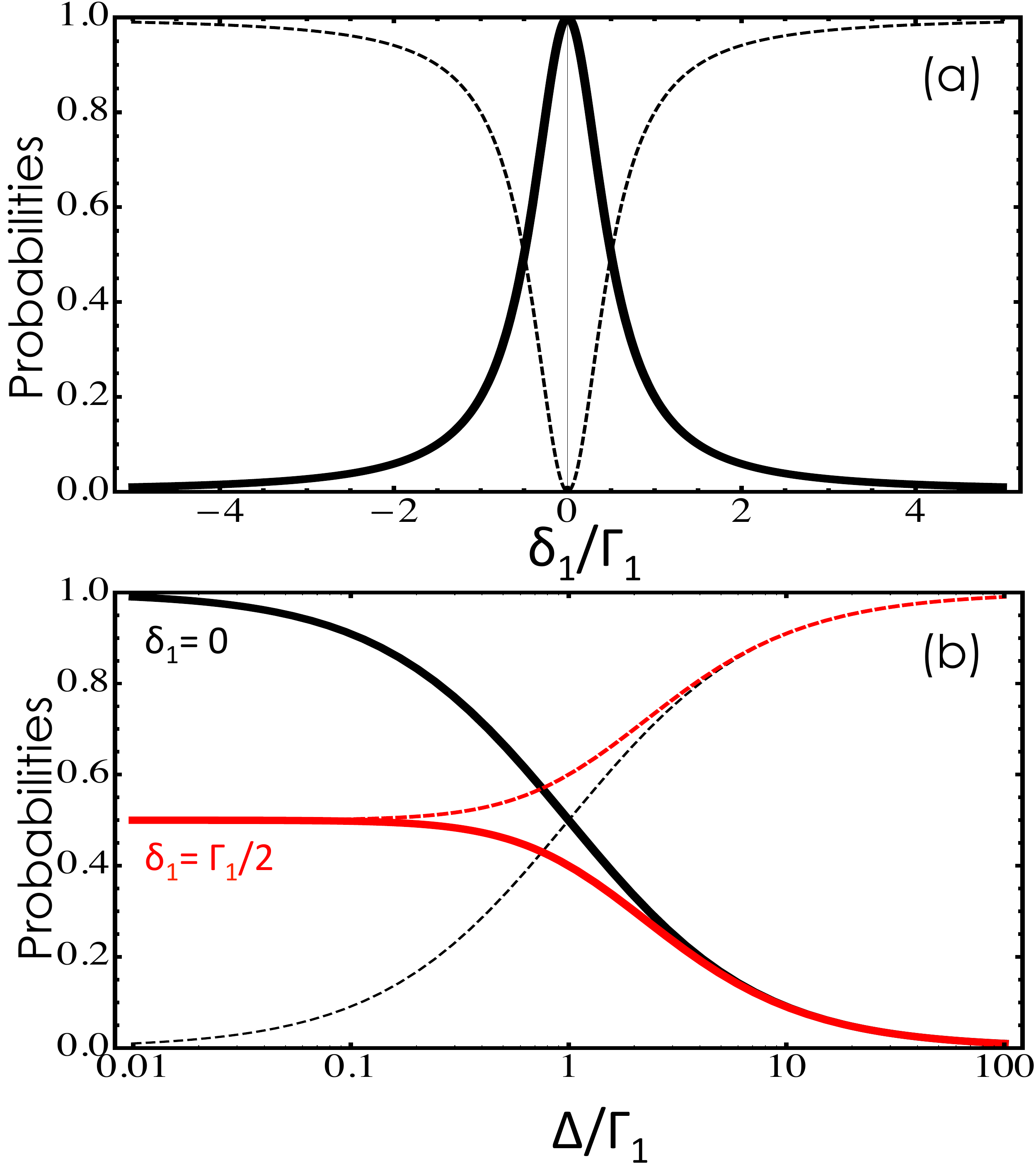}
\caption{(Color online)
{\bf Single quantum beamsplitter and its nonlinearity.}
(a) Photodetection probabilities $p^{(a)}_1$ (dashed) and $p^{(b)}_1$ (full) as a function of the detuning $\delta_1 = \omega_L - \omega_1$, in the monochromatic regime $\Delta = 0.001\Gamma_1$.
At resonance, TLS $1$ acts as a quantum mirror ($p^{(b)}_1=1$).
Off-resonance at $\delta_1 = \pm \Gamma_1/2$, the TLS acts as a balanced quantum beamsplitter ($p^{(a)}_1= p^{(b)}_1 = 1/2$).
(b) $p^{(a)}_1$ (dashed) and $p^{(b)}_1$ (full) as a function of the linewidth $\Delta$, for resonant (black) and off-resonance (red) regimes.
At $\Delta = \Gamma_1$ and $\delta_1 = 0$, the TLS also acts as a balanced beamsplitter.
} 
\label{fig2}
\end{figure}

\subsection{Quantum Mach-Zehnder interferometer}
\label{BS2}
Here we show our main result, namely, that two concatenated quantum beamsplitters can form the most elementary Mach-Zehnder interferometer.
We label them TLS $1$, as treated in Sec.(\ref{BS1}), and TLS $2$, with decay rate $\Gamma_2$ and transition frequency $\omega_2$.
The scattering on beamsplitter $2$ is obtained by assuming that the input state on $2$ equals the output state of $1$, solved in Sec.(\ref{BS1}).
More precisely, we assume that 
$\phi^{(a)}_2(z,0) = \phi^{(b)}_1(-z-ct_{\infty},\ t_\infty)$ and
$\phi^{(b)}_2(z,0) = \phi^{(a)}_1(-z+ct_{\infty},\ t_\infty)$.
Similarly to Sec.(\ref{BS1}), we consider large times $t_\infty \gg \Gamma_1^{-1} + \Delta^{-1}$.
This method is valid as long as the distance between the two TLSs is larger than $c\Delta^{-1}+c\Gamma^{-1}$, otherwise interference effects may qualitatively alter the dynamics (see, e.g., \cite{QFP,dudu,ff,prldiode}), going beyond the scope of the present paper.
We compute the output photodetection probabilities from beamsplitter $2$, $p^{(a)}_2$ and $p^{(b)}_2$.
As in the ideal Mach-Zehnder with classical beamsplitters \cite{haroche}, here we search for 
\beq
p^{(a)}_2 = 1 \ \ \mbox{and} \ \ p^{(b)}_2 = 0,
\label{MZ1}
\eeq
with
\beq
p^{(a)}_1 = p^{(b)}_1 = 1/2,
\label{MZ2}
\eeq
Eq.(\ref{MZ2}) consisting in the balanced condition.
We find analytical expressions for $p^{(a)}_2$ and $p^{(b)}_2$ as functions of the linewidth $\Delta$ and the detunings $\delta_1 = \omega_L - \omega_1$ and $\delta_{2} = \omega_L - \omega_2$ (see Appendix).
These are plotted in Figs.(\ref{fig3})-(\ref{fig5}).

Fig.(\ref{fig3}) shows $p^{(a)}_2$ (full line) and $p^{(b)}_2$ (dashed line) as a function of $\delta_1$, for identical beamsplitters $\delta_2 = \delta_1$, and $\Gamma_2 = \Gamma_1$, in the monochromatic regime $\Delta \ll \Gamma_1$.
Note that equally varying both detunings can be understood either as varying both TLS frequencies, $\omega_{1}$ and $\omega_2$, or as varying only the photon frequency $\omega_L$.
The properties of a Mach-Zehnder interferometer with classical beamsplitters, i.e., Eqs.(\ref{MZ1}) and (\ref{MZ2}), are obtained for the balanced detuning conditions  $\delta_{1} = \delta_2 = \pm \Gamma_1/2$.
\begin{figure}[!htb]
\centering
\includegraphics[width=1.0\linewidth]{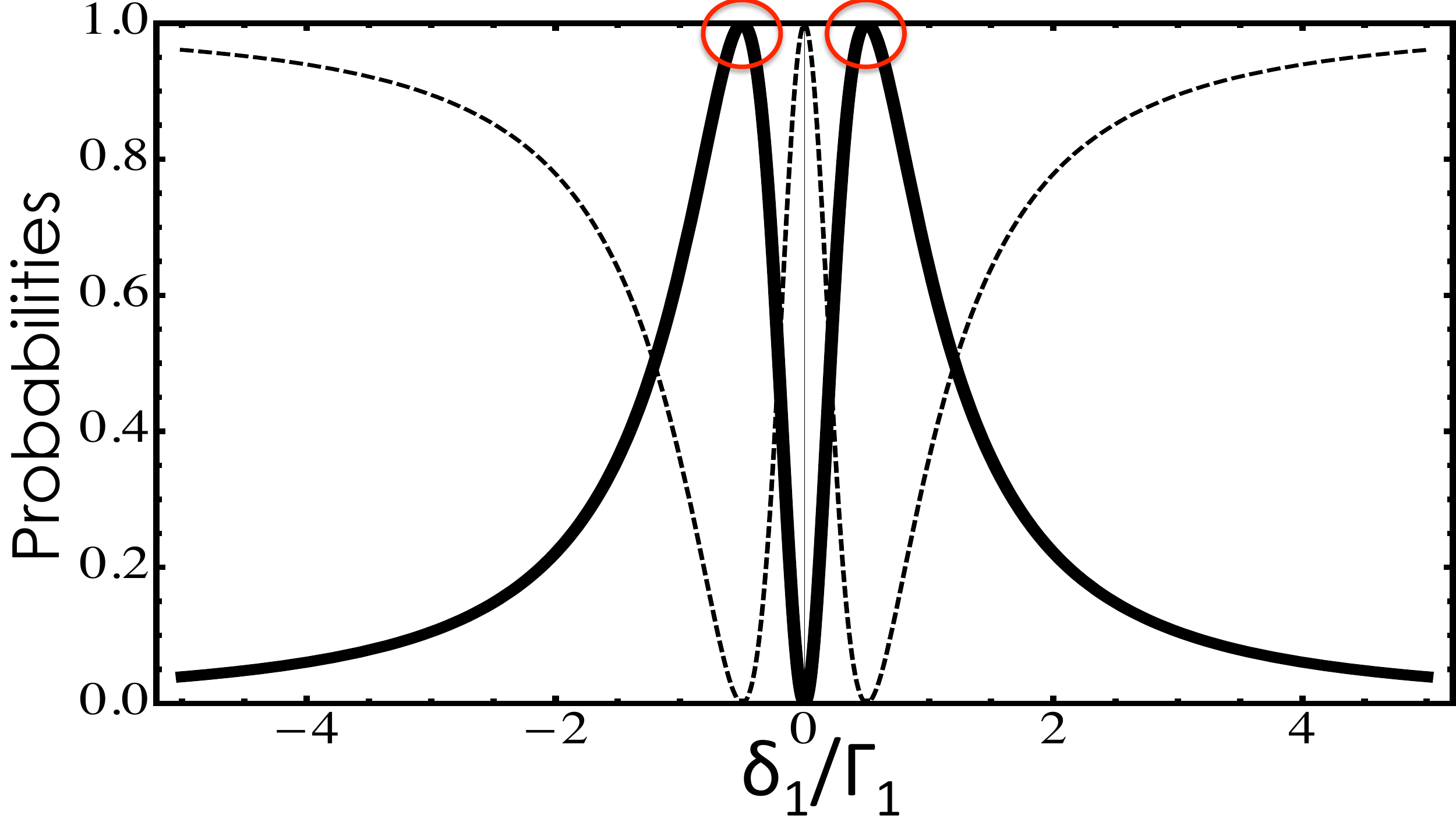}
\caption{(Color online)
{\bf 
Identical off-resonant TLSs in a QMZ behave as classical beamsplitters.}
Photodetection probabilities $p^{(a)}_2$ (full) and $p^{(b)}_2$ (dashed) as functions of the detuning $\delta_1 = \delta_2$ (equally varying both detunings can also be understood as varying the photon frequency $\omega_L$).
We set $\Gamma_2 = \Gamma_1$ and $\Delta =0.001 \Gamma_1$. 
Red circles highlight the equivalence with a Mach-Zehnder interferometer made of classical balanced beamsplitters, at $\delta_{1} = \pm \Gamma_1/2$.
} 
\label{fig3}
\end{figure}

Fig.(\ref{fig4}) shows $p^{(a)}_2$ (full line) and $p^{(b)}_2$ (dashed line) as a function of $\delta_2$, for fixed beamsplitter $1$, at $\delta_1 = \Gamma_1/2$.
Again we assume $\Gamma_2 = \Gamma_1$ and the monochromatic regime 
$\Delta \ll \Gamma_1$.
The peak at $\delta_2 = \delta_1 = \Gamma_1/2$ evidently reproduces the identical TLSs previously analyzed.
At $\delta_2 = -\Gamma_1/2$, a peak appears showing that $p^{(b)}_2 = 1$.
This breaks the analogy with classical balanced beamsplitters in a Mach-Zehnder interferometer of zero phase difference between the two paths. 
In this sense, $p^{(b)}_2 = 1$ provides a nonclassical signature for our QMZ.
This result simulates the presence of a $\pi$-phase shifter in one of the arms of the interferometer made of classical beamsplitters.
\begin{figure}[!htb]
\centering
\includegraphics[width=1.0\linewidth]{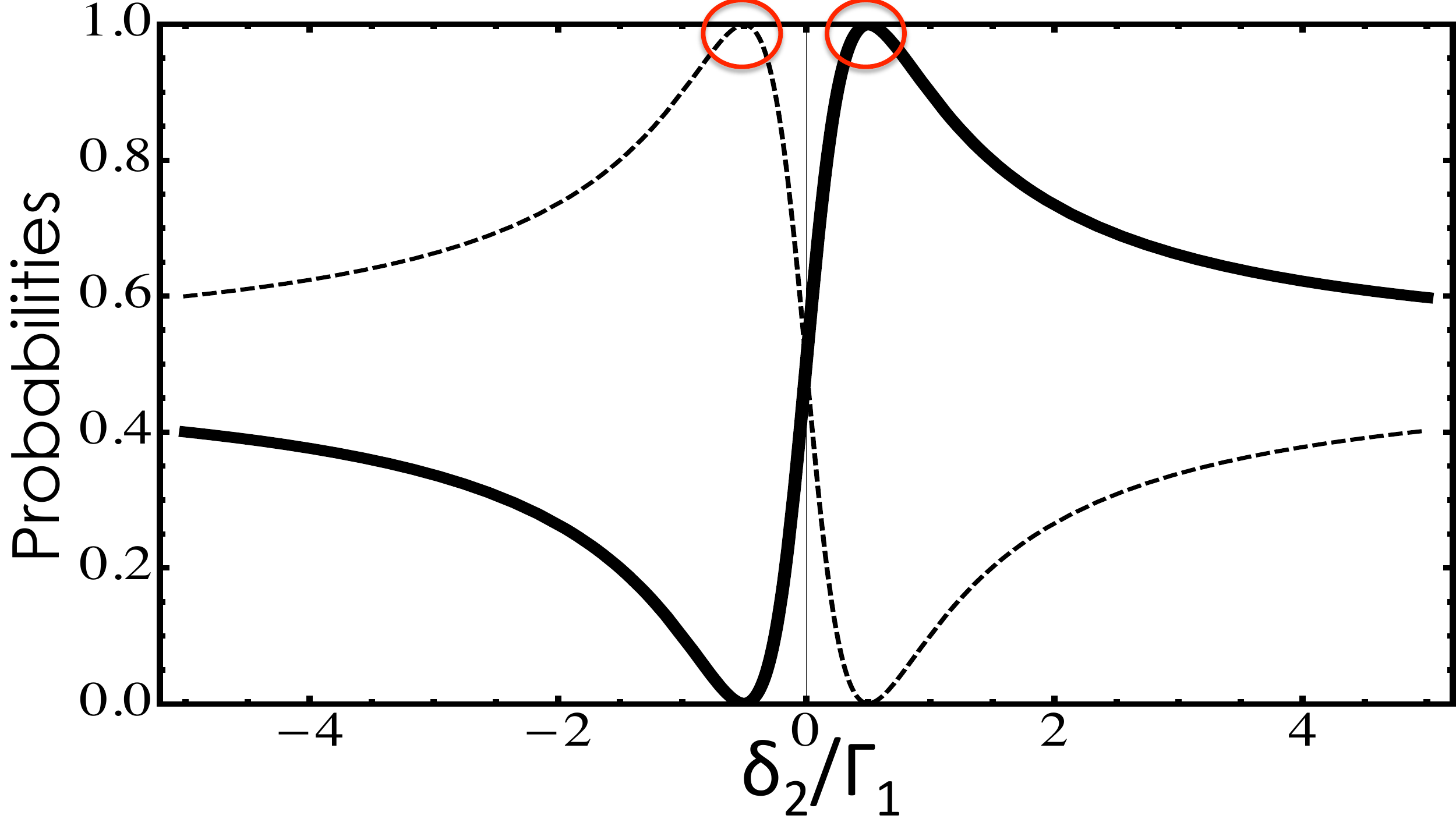}
\caption{(Color online)
{\bf Different TLSs in a QMZ present a nonclassical signature.}
Photodetection probabilities $p^{(a)}_2$ (full) and $p^{(b)}_2$ (dashed) as functions of the detuning $\delta_2$, for fixed $\delta_1 = \Gamma_1/2$.
We set $\Gamma_2 = \Gamma_1$ and $\Delta =0.001 \Gamma_1$.
The nonclassical signature takes place at $\delta_{2} = -\Gamma_1/2$ (circle on the left), in the sense that the photon enters the QMZ in channel $a_1$ and comes out from $b_2$ with certainty, even though there is zero phase difference between the two paths.
The classical behavior is found at the balanced condition of identical TLSs, $\delta_{2} = \delta_1 = \Gamma_1/2$ (circle on the right).
}
\label{fig4}
\end{figure}

Fig.(\ref{fig5}) presents the nonlinear properties of our QMZ with respect to the pulse linewidth.
We plot $p^{(a)}_2$ (full lines) and $p^{(b)}_2$ (dashed lines) as functions of $\Delta$.
Resonant TLSs are plotted in black ($\delta_2 = \delta_1 = 0$), off-resonance identical TLSs in red ($\delta_2 = \delta_1 = \Gamma_1/2$), and off-resonance different TLSs in blue ($\delta_2 = -\delta_1 = -\Gamma_1/2$).
We set $\Gamma_2 = \Gamma_1$.
We first analyze the resonant case (black).
Condition $\Delta = \Gamma_1$ implies a balanced beamsplitter (see Fig.\ref{fig2}(b)).
However, this very same condition does not allow for photon recombination into a single propagating mode.
Instead, we find that $p^{(a)}_2 = p^{(b)}_2 = 1/2$.
In this sense, the resonant QMZ presents a nonclassical signature that simulates a $\pi/2$-phase difference between the paths connecting classical beamsplitters in a Mach-Zehnder interferometer.
Similar behavior occurs for the off-resonant identical TLSs (red) around $\Delta \approx \Gamma_1$.
Remarkably, off-resonant different TLSs preserve unbalanced outputs for any $\Delta$, forming the most robust configuration to the deleterious effect of a finite linewidth.
Finally, it is worth reminding that the nonlinearity of the QMZ with respect to $\Delta$ arises from the fact that the balanced condition for quantum beamsplitters happen relatively close to the resonance with the TLS absorption frequency, mixing dispersive and absorptive contributions \cite{domokos,dv18}, in contrast to the case of classical beamsplitters.
\begin{figure}[!htb]
\centering
\includegraphics[width=1.0\linewidth]{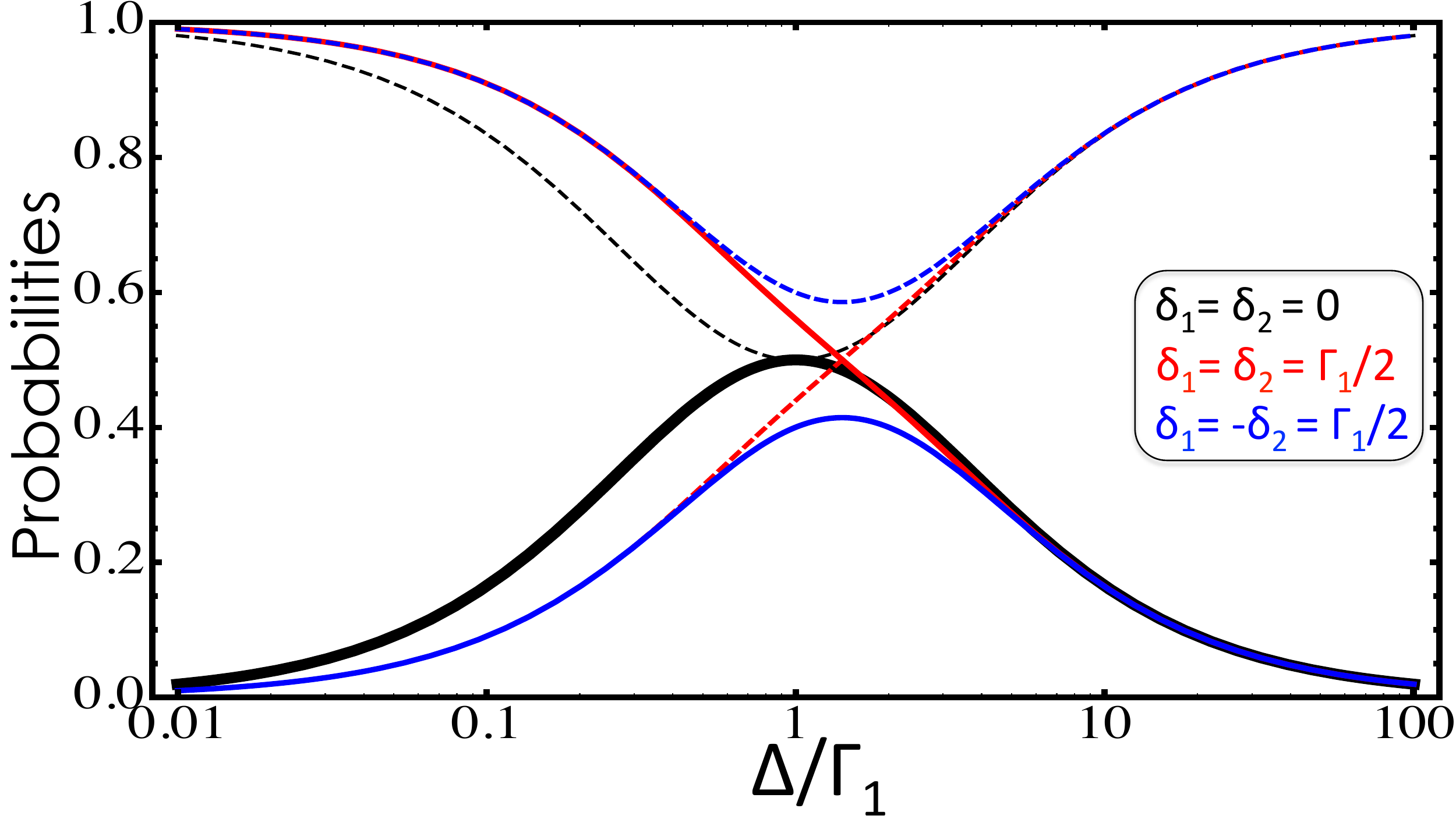}
\caption{(Color online)
{\bf QMZ nonlinearity.}
$p^{(a)}_2$ (full lines) and $p^{(b)}_2$ (dashed lines) as functions of the linewidth $\Delta$.
Resonant TLSs in black ($\delta_2 = \delta_1 = 0$), off-resonance identical TLSs in red ($\delta_2 = \delta_1 = \Gamma_1/2$), and off-resonance different TLSs in blue ($\delta_2 = -\delta_1 = -\Gamma_1/2$).
We set $\Gamma_2 = \Gamma_1$.
A nonclassical signature of the QMZ takes place for resonant TLSs (black) at the balanced condition $\Delta = \Gamma_1$, where a zero-phase difference QMZ simulates a classical-beamsplitter interferometer with a $\pi/2$-phase shifter in one path.
Remarkably, different TLSs (blue) maintain asymmetric outputs for any $\Delta$, forming the most robust configuration.
}
\label{fig5}
\end{figure}

\subsection{Transfer matrix for the monochromatic regime}
We show how to extend the transfer matrix from Ref.\cite{fan} to our QMZ.
%
Firstly, we analyze the transfer matrix for TLS $1$.
We define $M_1$ satisfying $\vec{\mbox{o}} = M_1 \vec{\mbox{i}}$, where $\vec{\mbox{o}} = [ a_{\mathrm{out}} \ \ b_{\mathrm{out}}]^T$ is the output modes vector and $\vec{\mbox{i}} = [ a_{\mathrm{in}} \ \ b_{\mathrm{in}}]^T$ is the input modes vector.
By recasting the results of \cite{fan} into the notations of this paper, we find that
\beq
M_1 = \frac{1}{1-i \lambda_1}
\begin{bmatrix}
1& i\lambda_1 \\  i\lambda_1& 1
\end{bmatrix},
\eeq
where $\lambda_1 = \Gamma_1/(2 \delta_1)$.
The balanced beamsplitter condition is satisfied at $\lambda_1=\pm 1$.

The transfer matrix for two concatenated TLSs is given by the product
$M_2 \sigma_x M_1$, where the Pauli matrix $\sigma_x$ is introduced to guarantee our convention that mode $a_{1,\mathrm{out}}$ becomes $b_{2,\mathrm{in}}$, as in Sec.(\ref{BS2}) and Fig.\ref{fig1}(a).

For identical TLSs, we have that $M_2 = M_1$.
For balanced beamsplitters $\lambda_1 = \lambda_2 = \pm 1$, we find that $M_2 \sigma_x M_1 = -\mathcal{I}$, where $\mathcal{I}$ is the identity matrix.
This means that a photon entering the QMZ in mode $a_{1,\mathrm{in}}$ splits and recombines at $a_{2,\mathrm{out}}$, as found in the full lines of Figs.(\ref{fig3}) and (\ref{fig4}).

For different TLSs, both working at the balanced condition, $\lambda_1=-\lambda_2=\pm 1$, we have that $M_2 = M_1^*$ and $M_2 \sigma_x M_1 = \sigma_x$.
This means that a photon entering the QMZ in mode $a_{1,\mathrm{in}}$ splits and recombines at $b_{2,\mathrm{out}}$, as found in the dashed line of Fig.(\ref{fig4}).

These results show that the transfer-matrix method is equivalent to the monochromatic regime ($\Delta \ll \Gamma_{1,2}$) of our QMZ, as presented in Sec.(\ref{BS2}).
Since no choice of initial pulse shape is required by this method, it generalizes the results obtained under the assumption of an initially exponential pulse, as in Eq.(\ref{initialpulse}).

\section{Conclusions}
We have analyzed how two quantum beamsplitters, consisting of TLSs in 1D waveguides, can form an elementary, fully-quantum Mach-Zehnder interferometer (QMZ) for single-photon pulses.
We demonstrate that our QMZ is equivalent to a Mach-Zehnder interferometer made of classical balanced beamsplitters, i.e., $p^{(a)}_2 = 1$ and $p^{(a)}_1 = 1/2$, for identical off-resonance TLSs, $\delta_1=\delta_2=\pm \Gamma_1/2$, in the monochromatic regime, $\Delta \ll \Gamma_1$.
We show a nonclassical signature of the QMZ in the monochromatic regime characterized by the recombination of the photon at mode 
$b_{2,\mathrm{out}}$, so that $p^{(a)}_2 = 0$, with balanced superposition $p^{(a)}_1 = 1/2$.
This takes place at off-resonance distinct TLS frequencies, $\delta_1 = -\delta_2 = \Gamma_1/2$.
In this case, the QMZ with zero phase difference between the two paths simulates a Mach-Zehnder of classical beamsplitters with a $\pi$-phase shifter in one of the paths.
We also show a nonclassical signature of our QMZ at finite pulse linewidths $\Delta \approx \Gamma_1$ for a resonant photon ($\delta_1 = \delta_2 = 0$).
In that case, we find balanced QMZ outputs, $p^{(a)}_2 = p^{(b)}_2 = 1/2$, simulating the effect of a $\pi/2$-phase shifter in a Mach-Zehnder of classical beamsplitters.
Finally, we have evidenced the robustness of different TLSs with respect to finite linewidths, as they maintain unbalanced outputs for all $\Delta$, in contrast to identical TLSs.
In the monochromatic regime, the quantum dynamical approach, necessary for characterizing finite linewidth nonlinearities, was shown to be equivalent to the transfer-matrix approach.
Our results open the path towards controllable elementary, fully-quantum interferometers that can be integrated in nanophotonic and superconducting circuit platforms.


\begin{acknowledgements}
N. A. acknowledges support from CAPES, Brazil. 
T. W. and D. V. acknowledge support from Instituto Nacional de Ci\^encia e Tecnologia -- Informa\c c\~ao Qu\^antica (INCT-IQ), CNPq Brazil.
\end{acknowledgements}

\

\section*{APPENDIX: Analytical Expressions}
We start by TLS $1$.
The output photodetection probability on channel $a_{1,\mathrm{out}}$ is 
$p^{(a)}_1 = 1-p^{(b)}_1$, 
where
\begin{eqnarray}
p^{(b)}_1 &=& \frac{\Gamma_1^2}{(\Gamma_1-\Delta)^2+(2\delta_1)^2}
\nonumber\\
&\times &\left( 
1 + \frac{\Delta}{\Gamma_1} - 
\frac{4\Delta (\Gamma_1+\Delta)}{(\Gamma_1 + \Delta)^2 + (2\delta_1)^2} \right)
\end{eqnarray}
is the output photodetection probability in channel $b_{1,\mathrm{out}}$.

The expressions characterizing the outputs of the Mach-Zehnder interferometer, i.e., of TLS $2$, are given below.
The photodetection probability on channel $a_{2,\mathrm{out}}$, as a function of $\Gamma_1$, $\Gamma_2$, $\Delta$, $\delta_1$ and $\delta_2$, reads
\beq
p^{(a)}_2 = \frac{1}{2}\left( \Lambda +2\ \Re[\mu] \right),
\eeq
where $\Re[.]$ stands for the real part. Here,
\beq
\Lambda = \frac{|B+K_1|^2}{\Delta} + \frac{|-B+K_2|^2}{\Gamma_1} + \frac{|K_1+K_2|^2}{\Gamma_2},
\eeq
and
\beq
\mu = \mu_{1}+\mu_{2}+\mu_{12},
\eeq
where
\beq
\mu_1 = \frac{(B+K_1)^*(-B+K_2)}{(\Delta + \Gamma_1)/2 -i \delta_1},
\eeq

\beq
\mu_2 = \frac{(B+K_1)^*(-K_1-K_2)}{(\Delta + \Gamma_2)/2 -i \delta_2},
\eeq

\beq
\mu_{12} = \frac{(-B+K_2)^*(-K_1-K_2)}{(\Gamma_1 + \Gamma_2)/2 -i (\delta_2-\delta_1)},
\eeq
and, finally,
\beq
B = \frac{-\Gamma_1 \sqrt{\Delta/2}}{(\Gamma_1-\Delta)/2-i\delta_1},
\eeq

\beq
K_1 = \frac{-(\Gamma_2/2)(\sqrt{2\Delta}+2B)}{(\Gamma_2-\Delta)/2-i\delta_2},
\eeq

\beq
K_2 = \frac{ \Gamma_2 B}{(\Gamma_2-\Gamma_1)/2-i(\delta_2-\delta_1)}.
\eeq
Quantum state normalization of the scattered photon implies that $p^{(b)}_2 = 1-p^{(a)}_2$.

\end{document}